\documentclass[twocolumn,showpacs,preprintnumbers,amsmath,amssymb]{revtex4}

\usepackage[pdftex]{graphicx}
\usepackage{dcolumn}
\usepackage{bm}

\begin{document}


\title{The Structure of Phonological Networks Across Multiple Languages}

\author{Samuel Arbesman}
\email{arbesman@hcp.med.harvard.edu}
 \affiliation{Department of Health Care Policy, Harvard Medical School, 180 Longwood Avenue, Boston, Massachusetts 02115}
\author{Steven H. Strogatz}
 \email{shs7@cornell.edu}
\affiliation{Theoretical and Applied Mechanics, Cornell University, Ithaca, New York 14853}

\author{Michael S. Vitevitch}%
\email{mvitevit@ku.edu}
\affiliation{Psychology, University of Kansas, Lawrence, Kansas 66045}

\date{\today}

\begin{abstract}
The network characteristics based on the phonological similarities in the lexicons of several languages were examined. These languages differed widely in their history and linguistic structure, but commonalities in the network characteristics were observed. These networks were also found to be different from other networks studied in the literature. The properties of these networks suggest explanations for various aspects of linguistic processing and hint at deeper organization within human language.
\end{abstract}

\maketitle

\section{Introduction}

The results of numerous graph-theoretic analyses suggest that a number of principles may influence the emergent structures found in a wide variety of complex systems, including information, social, technological, and biological networks \cite{Strogatz01,Albert02,Newman03}. These unifying characteristics include small-world properties, distinct community structure, and scale-free distributions of the network connectivity.

Many aspects of language can be examined from a network perspective as well. Numerous studies have been conducted on semantic networks, where relationships in meaning have been made between words. These are often based on thesauri, word-associations in corpori or from academic databases \cite{Ferrer-i-Cancho01,Motter02}. In addition, linguistic networks have been made from orthographic similarities of words (how words are spelled) \cite{KelloBeltz}. Lastly, language can be viewed from the sounds of words (their phonological structure), where words that sound similar are neighbors. Although previous experiments have examined  small portions of phonological networks (nearest neighbors of words) in the context of psycholinguistic theories of spoken word recognition \cite{Luce98}, the first graph-theoretic analysis of an entire language network only appeared more recently \cite{Vitevitch08}.

\begin{figure}
\includegraphics[width=3 in]{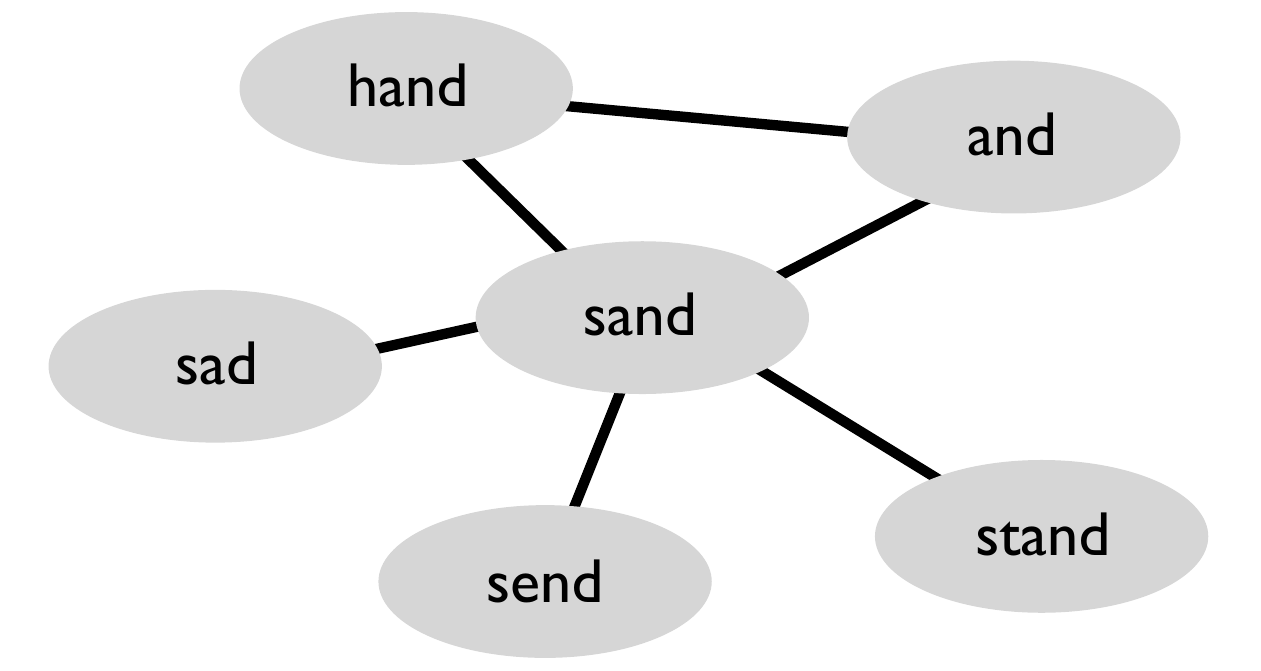}
\caption{\label{fig_1_new} A phonological network for five English words.}
\end{figure}

In these phonological networks, words in a language are represented as vertices or nodes, and an edge is placed between them if the words sound similar to each other (differing only by a single phoneme, or sound segment). For example, as shown in Figure \ref{fig_1_new}, vertices representing the words \textit{hand}, \textit{send}, \textit{sad}, \textit{and}, and \textit{stand} would all have edges connecting them to the vertex for the word \textit{sand}. These phonological networks are especially intriguing to examine because psycholinguistic studies suggest that several characteristics of the network influence cognitive processing, such as word recognition and retrieval \cite{Steyvers05,Vitevitch08}.  

In examining English, Vitevitch \cite{Vitevitch08} found that its phonological network had a small giant component (the largest connected portion of the graph), with many other smaller components ("islands"). This property is distinct from other complex networks observed in the literature. In addition, the degree distribution (the distribution of the number of edges per node) was not well modeled by a scale-free distribution, or a power law.

Here, we wanted to explore the generality of these results, by doing the first comparative study of multiple languages, using phonological networks. We examined some of the properties looked at by Vitevitch in English, as well as a number of others, and found that phonological networks all have certain properties distinct from other types of complex networks (such as biological and social networks).

\section{Methods}

The network structure of selected languages was examined to determine the generality of the network characteristics previously observed in English \cite{Vitevitch08}. In addition to English, the following languages were examined: Spanish, Mandarin, Hawaiian, and Basque (see Table \ref{table_1}). Similar network characteristics across a variety of languages might hint toward principles that are common to all languages, whereas differences in network measures might provide a quantitative way to describe and categorize the languages of the world.
	
English is an Indo-European language from the Germanic branch, whereas Spanish comes from the Romance branch of the Indo-European family of languages. Mandarin, a Sino-Tibetan language, differs from English, Spanish, Hawaiian and Basque in that it also uses tones to convey word meanings (e.g., "fan" with a high level tone means sail, with a rising tone means trouble, with a dipping tone means turn, and with a falling tone means rice). Tone was not included in the phonological transcriptions, however. Hawaiian is an Austronesian language with a phoneme inventory (the number of consonants and vowels in the language) that is smaller than those found in English, Spanish, Mandarin, and Basque. Finally, Basque (or Euskara) is a linguistic isolate, meaning that it is not (or has not yet been identified as) a member of a given language family. Additional differences, such as those in morphology, exist among the languages that were selected for the present network analyses.
	
The phonological networks were constructed from a variety of sources. The English network contained the words from the Merriam-Webster Pocket Dictionary from 1964; this database has been used extensively in psycholinguistic studies \cite{Luce98}. The Hawaiian network was created in a similar manner using a Hawaiian Dictionary \cite{Judd80}. The words from the Spanish network consisted of the words in the LEXESP database \cite{Sebastian00}, a large Spanish language corpus. The words in the Basque network were obtained in a manner similar to the words in the Spanish network \cite{Perea06}. The Mandarin network uses the words from a database compiled in \cite{Huang97}. 

\section{Results}

\begingroup
\squeezetable
\begin{table*}
\caption{Summary information of phonological networks in several languages. GC stands for Giant Component and RN stands for Random Network.}
\centering
\begin{tabular}{l c c c c c}
 & \em{English} & \em{Spanish} & \em{Mandarin} & \em{Hawaiian} & \em{Basque}\\
	\hline
Network Size (number of words) & 19,323 & 122,066 & 30,086 & 2,578 & 99,321 \\ 
Giant Component Size (percentage) & 6,498 (0.34) & 44,833 (0.37) & 19,712 (0.66) & 1,406 (0.55) & 35,173 (0.35) \\ 
Assortative Mixing by Degree ($r$) & 0.657 & 0.762 & 0.654& 0.556& 0.719 \\ 
Average Shortest Path Length & 2.7& 4.3 & 6.5 & 3.2 & 4.4 \\
Average Shortest Path Length (GC) & 6.1 & 10.3 & 10.1 & 5.5 & 10.4 \\ 
Average Shortest Path Length of RN (using GC) & 5.8 & 9.9 & 7.3 & 5.8 & 11.4 \\ 
Clustering Coefficient & 0.284 & 0.191 & 0.383 & 0.241 & 0.206 \\ 
Clustering Coefficient of RN & 8.35e-5 & 1.17e-5 & 8.55e-5 & 7.40e-4 & 1.21e-5 \\ 
Transitivity & 0.313 & 0.250 & 0.404 & 0.260 & 0.232 \\ 
Ratio of Edges to Vertices & 1.61 & 1.43 & 2.57 & 1.91 & 1.21 \\ 
Ratio of Edges to Vertices (GC) & 4.55 & 2.95 & 3.88 & 3.44 & 2.50 \\ 
\end{tabular}
\label{table_1}
\end{table*}
\endgroup

\subsection{Unique Characteristics of the Giant Component}

\subsubsection{Giant Component Size}

The giant component sizes of the language networks were much smaller compared to other network structures discussed in the literature. Typically, the giant component contains approximately $80-90\%$ of the vertices \cite{Newman01}. However, in the present networks, the proportion of vertices in the giant component was much smaller, with some networks having less than $50\%$ of the vertices in the giant component. The proportion of vertices in the giant components for comparably sized random networks, containing $70-80\%$ of the vertices, are also larger than the values for the language networks \cite{Callaway01}. This difference in giant component size suggests that these phonological networks may be more robust to node removal due to more tightly connected components, and indicates the prevalence of smaller components in the networks. 

\subsubsection{Robustness to Vertex Removal}

\begin{figure}
\includegraphics[width=3 in]{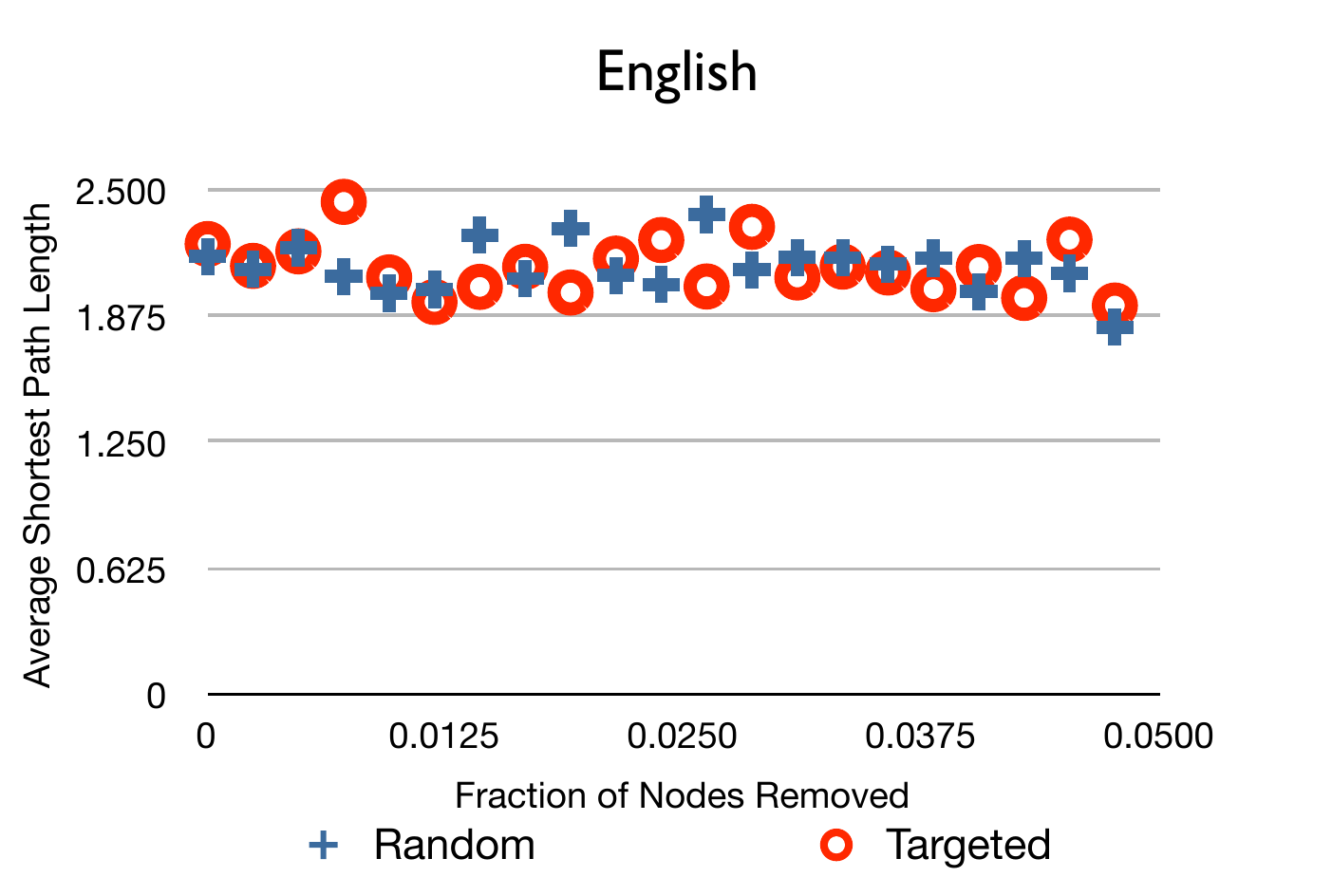}
\caption{\label{fig_1} An example run of node removal in English, either random or in a targeted fashion (in order by degree). Up to $5\%$ of the nodes were removed, and all languages showed similar patterns to the above results. In addition, when the simulations were done only for the giant component, a similar constant, though elevated, value of the average shortest path length was found.}
\end{figure}

To evaluate the robustness of the networks, vertices were removed in two ways: at random, and in decreasing order by degree (number of edges connected to a vertex). These results are shown in Figure \ref{fig_1}. In scale-free networks, when vertices are randomly removed the mean shortest path length remains constant, whereas when vertices are removed in order of degree, the mean shortest path length increases dramatically \cite{Newman03}. In the language networks, however, both methods of node removal resulted in little to no change in the mean shortest path lengths. The shortest path lengths were calculated using a sampling technique where $1,000$ nodes were chosen at random. Then, the distance to all other nodes (if part of the same component) were obtained and these paths lengths were then all averaged, to give an estimate of the shortest path length. This sped up the calculations considerably. The extraordinary amount of robustness observed based on these common methods of node removal does seem intriguing and merits further examination.

\subsubsection{Assortative Mixing}

In addition, we examined the assortative mixing by degree of the language networks, which is a measure of the correlation of degree between neighboring nodes. As seen in Table \ref{table_1}, all of the language networks had large and positive correlations of the degrees of connected vertices, indicating that high degree vertices tended to be connected to each other. Newman \cite{Newman02} discussed how networks with assortative mixing by degree are more robust to vertex removal and percolate more easily (i.e., diseases or information spread easily) than networks with disassortative mixing. The high assortative mixing observed in the phonological networks is distinct from other types of networks: biological and technological networks often are disassortatively mixed, and social networks, which display assortative mixing, still have lower values of assortative mixing. Typical measures of assortativity for social networks are $0.1-0.3$, and biological and technological networks are $-0.1$ to $-0.2$  \cite{Newman02}. On the other hand, phonological networks can be higher than $0.7.$

High assortative mixing not only suggests robustness in the phonological networks, and highlights the resilience of lexical processing in the face of injury to the language related areas of the brain (i.e., stroke), but it also has implications for the searchability of the phonological networks under intact conditions \cite{Watts02}. This feature of the phonological network may contribute to the high rates of accuracy with which words are retrieved from the mental lexicon; one study estimated that healthy adult speakers make an error between $0.1-0.2\%$ of the time they speak \cite{Garnham81}. Lexical processing might proceed more slowly and errors in word retrieval might be more common if the phonological networks did not have such a robust structure. The phonological networks of patients with aphasia or other neurogenic disorders that disrupt language processing could be used to test this hypothesis.

\subsection{Small-world properties}

Although the languages differ in their history and linguistic characteristics, they all share a number of similarities in their network structure. An important commonality across the languages is that they all have the properties of a small-world network \cite{Watts98}, that is, a high clustering coefficient and short vertex-to-vertex distance. The clustering coefficient can be calculated for each node (the average value of which is reported above in Table \ref{table_1}), and is the fraction of neighbors of a given node that are neighbors with each other. It is also known as network density. The vertex-to-vertex distance, also known as the shortest path length, is the shortest number of hops in a network to go from one node to another. Since these networks have many components, the shortest path length from one node to another is only calculated for nodes that are in the same component \cite{Newman03}. In addition, the mean shortest path length was calculated just within the giant component of each language.

As seen in Table \ref{table_1}, the values for the clustering coefficient are many orders of magnitude larger than what would be expected from a comparably sized random network\textemdash a network with the same number of nodes and edges\textemdash which can be calculated analytically \cite{Watts98}. The values of the clustering coefficient are also comparable to a similar measure referred to as transitivity, which is a more global measure of clustering \cite{Newman03}.

On the other hand, the mean shortest path length of the language networkÕs giant component, calculated using a random sample of $1,000$ nodes, was similar to the mean shortest path length for comparably sized random networks, and significantly shorter than the overall number of nodes in the network, as seen in Table \ref{table_1} \cite{Watts98}. The statistics of the giant component were used for comparable random networks, because the overall ratio of edges to nodes is far lower than within the giant component, due to the large number of islands in the networks.

Since a small world structure is often a prerequisite for rapid search, and it is well-known that lexical retrieval processes are rapid and robust, it would be logical that the networks might be optimally structured for search. A clear future research direction is the examination of these networks for the properties, such as those discussed in Kleinberg \cite{Kleinberg00}, that allow for rapid and robust search.

However, it must be noted that, unlike in social networks, where it is clear what a distance of three friends is, for example, it is not entirely clear what the qualitative difference is between a distance of 5 and 6 within phonological networks. This is important when looking at the average shortest path lengths of the giant components of the different language networks. For instance, is it relevant that this value for Mandarin (10.1) is twice that of Hawaiian (5.5)? While it is likely that this number is most relevant relative to the size of the entire network (they are all orders of magnitude smaller than the size of the lexica examined), these differences might hint at more significant distinctions between the languages examined.

The common occurrence of the small world property  in networks observed may suggest that it is less a relevant property of language than simply an indicator that language is a fairly organic, unplanned construct. It is interesting, however, that the path length within a network appears to be an important property for language processing. A recent study \cite{Yarkoni08} demonstrated that a measure related to path length in a phonological network (i.e., the minimum number of substitution, insertion, or deletion operations required to turn one word into another) influenced pronunciation times in visual word recognition tasks. Therefore, the relevance of different average path length across languages warrants further investigation.

\subsection{Degree Distribution}

The degree distributions of scale-free networks obey a power law function,  $P(z) \sim z^{-\alpha}$. In contrast to many observed networks, we find that the language networks deviate from this behavior. Instead, they are reasonably fit to truncated power laws, similar to scientific co-authorship networks \cite{Newman01}, as seen in Table \ref{table_2}. A truncated power law, or a power law with an exponential cutoff, is defined as follows:
\begin{equation}
P(z) \sim z^{-\alpha} e^{-z/z_c}
\end{equation}

Table \ref{table_2} shows the parameters of the best fit of a truncated power law for the degree distribution of each language, as calculated by the methods found in Clauset et al. \cite{Clauset07}. All fits had p-values of less than $10^{-10}$, in terms of the probability that they were better fit by a truncated power law than a traditional power law. In addition, as can be seen, Mandarin's fit is essentially an exponential distribution, with no power-law portion.

\begin{table}
\caption{Languages and best fit parameters for a truncated power law. All fits had p-values of less than $10^{-10}$.}
\centering
\begin{tabular}{l c c}
Language & Exponent ($\alpha$) & Cutoff ($z_c$) \\
	\hline
\em{English} & 0.826 & 16.14 \\ 
\em{Spanish} & 0.815 & 7.06 \\ 
\em{Mandarin} & -1.0 & 3.69 \\ 
\em{Hawaiian} & 0.270 & 7.34  \\
\em{Basque} & 0.575 & 4.56 \\ 
\end{tabular}
\label{table_2}
\end{table}

Amaral et al. \cite{Amaral00} found that if there is a constraint associated with the attachment of  a new vertex (i.e., the vertex may only be able to accommodate a fixed number of edges), then a power law degree distribution, like that in the scale-free model proposed by Barab‡si and Albert \cite{Barabasi99}, is not likely to be observed. In the language networks, a variety of constraints on word formation are present, such as the number of phonemes in the inventory of the language, the sequential arrangement of phonemes in words, the length of words, and the extent to which the language relies on morphemes (the smallest meaningful unit). All of these constraints limit the number of words that might be phonologically similar. Therefore, a truncated power law or similar distributions that decay faster than a traditional power law are reasonable as fits for the degree distributions in phonological networks.

\begin{figure}
\includegraphics[width=3 in]{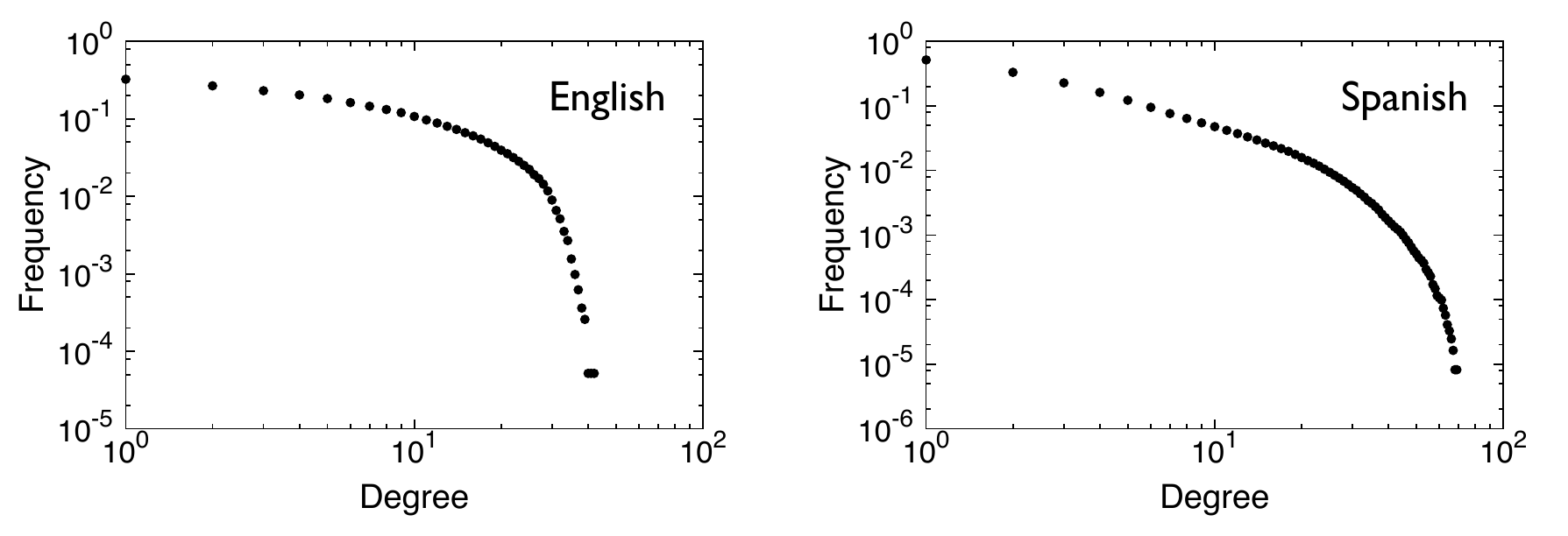}
\caption{\label{fig_2} The degree distributions of two of the language networks (English and Spanish), on a log-log scale. The final point for each distribution was not plotted, for legibility.}
\end{figure}

\section{Conclusion}

The phonological networks of a variety of languages show a unique structure not found in other complex networks described in the literature. Despite coming from a diverse range of language families the networks all exhibited a common set of properties. Notably, the degree distribution is found to lie somewhere between a power law and an exponential distribution.

Furthermore, a small-world structure was observed, in conjunction with the distinguishing characteristic of the giant components as far smaller than typically observed. The small sizes of the giant component together with the strong assortative mixing by degree and the robustness of the network to the removal of vertices is suggestive into the resilience of language processing in the brain, although further study is necessary.

Together, these observed characteristics hint at some deeper organization within language. Despite surface differences among languages, there are important commonalities that have implications for the processing of language in humans. The intriguing characteristics of these networks merit further investigation from network scientists as well as psycholinguistic researchers.

\section{Acknowledgments}

Research supported in part by National Science Foundation grant DMS-0412757 to S.H.S, and in part by grants from the National Institutes of Health to the University of Kansas through the Schiefelbusch Institute for Life Span Studies (National Institute on Deafness and Other Communication Disorders (NIDCD) R01 DC 006472), the Mental Retardation and Developmental Disabilities Research Center (National Institute of Child Health and Human Development P30 HD002528), and the Center for Biobehavioral Neurosciences in Communication Disorders (NIDCD P30 DC005803) to M.S.V. 

\bibliography{language}

\end{document}